# Model estimates for contribution of natural and anthropogenic CO$_2$ and CH$_4$ emissions into the atmosphere from the territory of Russia, China, USA and Canada to global climate changes in the 21st century

Denisov S.N.[1,2], Eliseev A.V.[1,2,3], Mokhov I.I.[1,2]
[1]A.M. Obukhov Institute of Atmospheric Physics RAS
[2]Lomonosov Moscow State University
[3]Kazan Federal University

denisov@ifaran.ru

**Abstract**

The contribution of anthropogenic and natural greenhouse gases to the atmosphere from the territory of Russia, China, USA and Canada to global climate change under different scenarios of anthropogenic emissions in the 21st century has been assessed. It is shown that the consideration of the changes in climate conditions can affect the impact indicators of greenhouse gas emissions on the climate system, especially over long time horizons. In making decisions, it is necessary to take into account that the role of natural fluxes of greenhouse gases into the atmosphere from the terrestrial ecosystems can change. For all the countries considered, the uptake of CO$_2$ by terrestrial ecosystems under all scenarios of anthropogenic impact begins to decrease in the second half of the 21st century, so its stabilizing effect may gradually lose importance. At the same time, methane emissions in all of the considered regions are increasing significantly. The net effect of these greenhouse gas natural fluxes in some cases can even lead to warming acceleration by the end of the 21$^{st}$ century.

**Introduction**

A detailed and comprehensive analysis of the impact of changes in the carbon cycle in the Earth's climate system requires, among other things, adequate consideration of the carbon balance of boreal forests, wetlands and other ecosystems (Climate Change 2013; SRCCL). This is especially relevant in connection with the Paris Agreement (2015) of the United Nations Framework Convention on Climate Change, concerning the problems of reducing greenhouse gas emissions and related adaptation at ational levels (Mokhov 2020).

Various indicators can be used to quantify the relative and absolute contribution to climate change of emissions of different greenhouse gases, as well as emissions from different regions, countries or individual sources. They serve to estimate different effects (such as changes in temperature or sea level) over different time horizons. The climatic effect of emissions can be estimated for a specific moment or integrated over a given time interval. The most common indicators are based on the radiative forcing (RF) (Shine et al. 2005; Karol' et al. 2011), which is used to compare the contribution to changes in global mean surface temperature of various factors affecting the Earth's radiative balance.

The UN Framework Convention on Climate Change, the Kyoto Protocol, and the Paris Agreement use 100-year global warming potentials (GWP), calculated as the integral of radiative forcing on a 100-year time horizon, to determine the relative role of anthropogenic emissions of various greenhouse gases. At the same time, targets for climate policy are usually formulated as some given temperature thresholds to be avoided, e.g., the 2°C global temperature rise limit outlined in the Copenhagen Accord (2009) or even 1.5ºC as targeted in the Intergovernmental Panel on Climate Changes Special Report on Global Warming of 1.5°C. Such goals are conceptually incompatible with a metric based on cumulative radiative forcing (Smith et al. 2012). The most used of the alternative metrics is the global temperature change potential (GTP,

(Shine et al. 2005; Shine et al. 2007)). The GTP indicator reflects changes in global temperature at a selected time interval after a pulse release of a selected gas relative to changes resulting from a similar CO2 release, and thus accounts for climate response along with radiative efficiency and atmospheric lifetime of the gas. This approach is directly consistent with the current goals of climate agreements.

In this paper estimates of anthropogenic and natural fluxes of $CO_2$ and $CH_4$ from the territory of different countries in the Northern Hemisphere (Russia, China, USA, Canada) in the 21st century were obtained using the climate model of the Obukhov Institute of Atmospheric Physics RAS (IAP RAS CM) (Mokhov and Eliseev 2012; Eliseev et al. 2014; Denisov et al. 2015) (see also (Mokhov et al. 2002; Mokhov et al. 2005; Eliseev and Mokhov 2011)). Cumulative influence of the $CO_2$ and $CH_4$ anthropogenic and natural fluxes to the surface air temperature changes since 1990 is estimated using the cumulative temperature potential CT based on GTP, which was modified to account for changing background conditions.

**Methods**

**Model and Setup of Numerical Experiments**

The IAP RAS CM belongs to the class of the global climate models of intermediate complexity (Claussen et al. 2002; Zickfeld et al. 2013; Eby et al. 2013; MacDougal et al. 2020). A specific feature of the model is that the large-scale atmospheric and oceanic dynamics (with a scale exceeding the synoptic) are described explicitly, whereas the synoptic processes are parameterized. The latter makes it possible to substantially reduce the time required for model simulations. The model contains modules of the carbon cycle including partly interactive methane cycle and a module for calculation of emissions from deforestation and from natural fires (Eliseev et al. 2008; Denisov et al. 2013).

Using the IAP RAS CM, we performed numerical experiments for 1765–2100 with scenarios of anthropogenic impacts on climate due to changes in the content of greenhouse gases in the atmosphere, tropospheric and stratospheric volcanic sulfate aerosols, changes in the total solar irradiance, and changes in the area of agricultural lands. For 1700–2005, these forcings were given in accordance with the "Historical Simulations" of the CMIP5 project (http://www.iiasa.ac.at/web-apps/tnt/RcpDb). For 2006–2100, the anthropogenic forcings were prescribed in accordance with the anthropogenic impact scenarios RCP 2.6 4.5 6.0 and 8.5 of the CMIP5 project. The numerical experiments described in this study are similar to those conducted earlier with the IAP RAS CM (Denisov et al. 2015; Denisov et al. 2019).

**Cumulative temperature potential**

Global Temperature Potential of gas x is the ratio of its absolute potential to absolute potential of $CO_2$:

$$GTP_x(H) = \frac{P_x^{(a)}}{P_{CO_2}^{(a)}}, \tag{1}$$

where the Absolute Global Temperature change Potential ($P^{(a)}$) is the change in global mean surface temperature at time H in response to a 1 kg pulse emission of gas x at time t = 0. It is often written as a convolution of the radiative forcing with the climatic response kernel $R_T$:

$$P_x^{(a)}(H) = \int_0^H RF_x(t) R_T(H-t) dt, \tag{2}$$

where $RF_x$ is the radiative forcing due to a pulse emission of a gas x, and $R_T$ is the temporally displaced climate response to a unit forcing. $RF_x$, can be written as the product of gas x radiative efficiency, $A_x$, and the perturbation in its burden, $IRF_x$ (see Supplementary material).

It should be noted that both $A_x$ and $IRF_x$, and hence $P^{(a)}$, are determined for a pulse gas emission under constant background conditions, while the assessment of impact of emission scenarios under changing conditions in the 21st century is required.

For changing background conditions $P^{(a)}$ can be rewritten as sum of integrals for each year:

$$P_x^{(a)*}(T_0, T_H) = \sum_{k=T_0+1}^{T_H} \int_{k-1}^{k} RF_{x,k}(t) R_T(T_H - T_0 - t) dt, \quad (3)$$

where $T_0$ is the year of emission and $T_H = T_0 + H$. $RF_{x,k}$ can be achieved assuming that all required parameters are constant for each particular year k, but able to change from year to year (see Supplementary material).

For sustained emissions started at time $T_0$, the cumulative effect of the source of gas x at time $T_H$ can be written as cumulative temperature potential:

$$CT_x(T_0, T_H) = \sum_{t=T_0}^{T_H-1} E_x(t) P_x^{(a)*}(t, T_H). \quad (4)$$

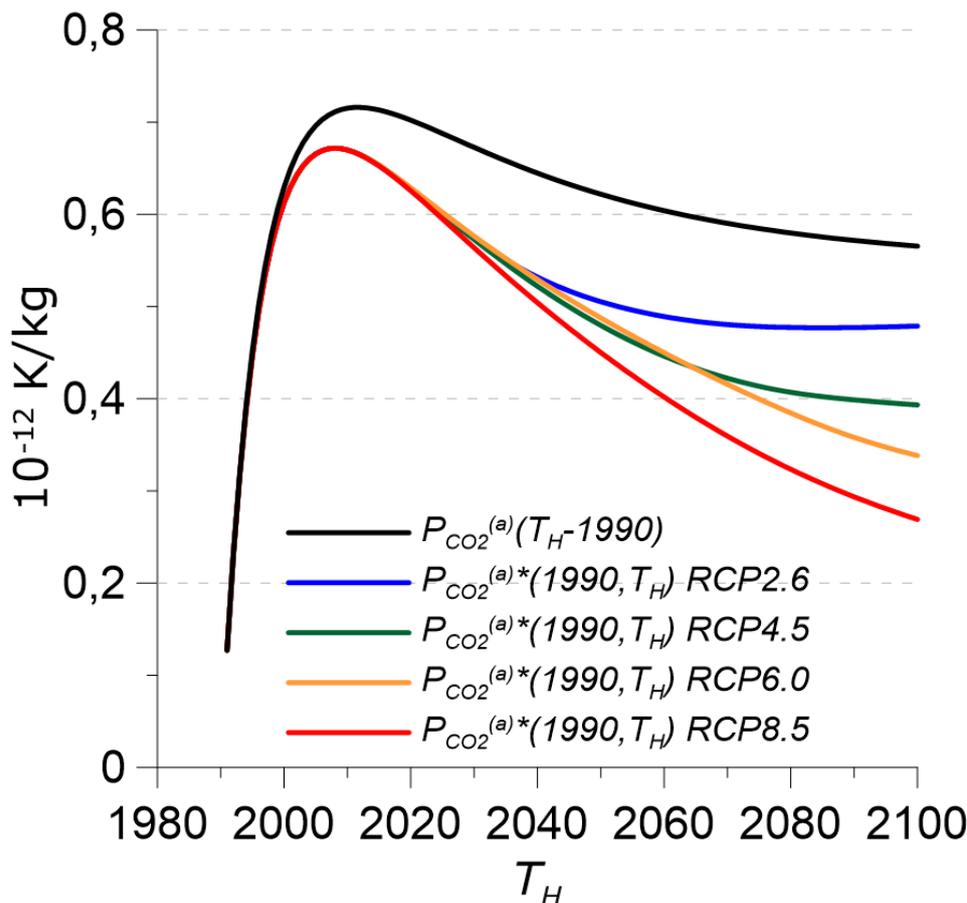

Fig. 1 Absolute potential of global temperature change ($P^{(a)}$ and $P^{(a)*}$) for pulse $CO_2$ emission in 1990 on the [1990,$T_H$] horizon

Figure 1 shows the absolute potentials $P^{(a)}$ and $P^{(a)*}$ of carbon dioxide for the time interval from year 1990 to year $T_H$ (thereafter denoted as [1990;$T_H$]) obtained from IAP RAS CM simulations. As mentioned above, the RF for $P^{(a)}$ is calculated assuming that background conditions at time $T_0$ remain unchanged throughout the whole interval $[T_0;T_H]$, whereas for $P^{(a)}*$ the change in RF due to changes in background conditions is taken into account. Therefore, for time intervals longer than 10-15 years, $P^{(a)}*$ and $P^{(a)}$ may differ significantly. For the most aggressive anthropogenic scenario RCP 8.5 (with the strongest change in background conditions) the $P^{(a)}$ is more than two-fold higher than $P^{(a)}*$ for carbon dioxide emitted in 1990 on the 100-year time interval.

The potentials $P^{(a)}$ and $P^{(a)}*$ for methane calculated for the interval $[1990;T_H]$ (Fig. 2a) differ significantly weaker for all scenarios considered. For RCP 8.5, the maximum discrepancy reach 20% for a time horizon of about 80 years, for other scenarios it does not exceed 6%. Nevertheless, if $P^{(a)}*$ is used to calculate the relative GTP potential of methane (Fig. 2b) instead of $P^{(a)}$, its values for long periods can be 2-2.5 times higher. Thus, the 100-year GTP of methane emitted in 1990, calculated using $P^{(a)}$, equals 4.1 (which corresponds to IPCC AR5), and taking into account changes in background conditions it can be 5-10, depending on the scenario of anthropogenic forcing.

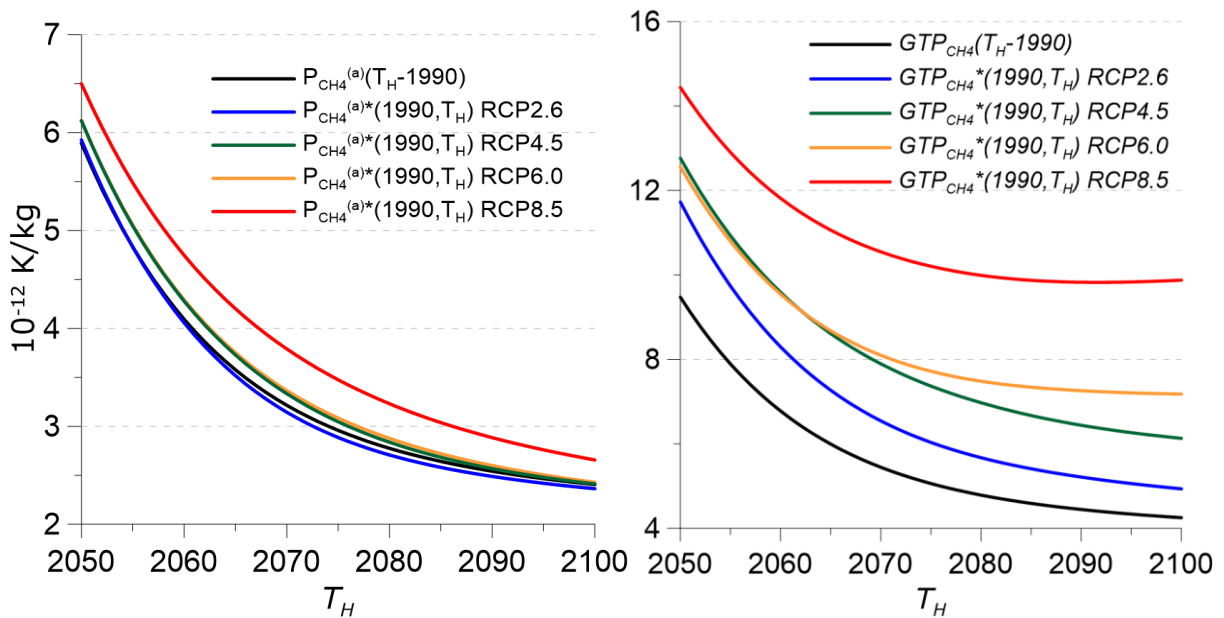

Fig. 2. Absolute global temperature change potential for pulse $CH_4$ emission in 1990 at the $[1990,T_H]$ horizon (a) and its relative GTP potential (b).

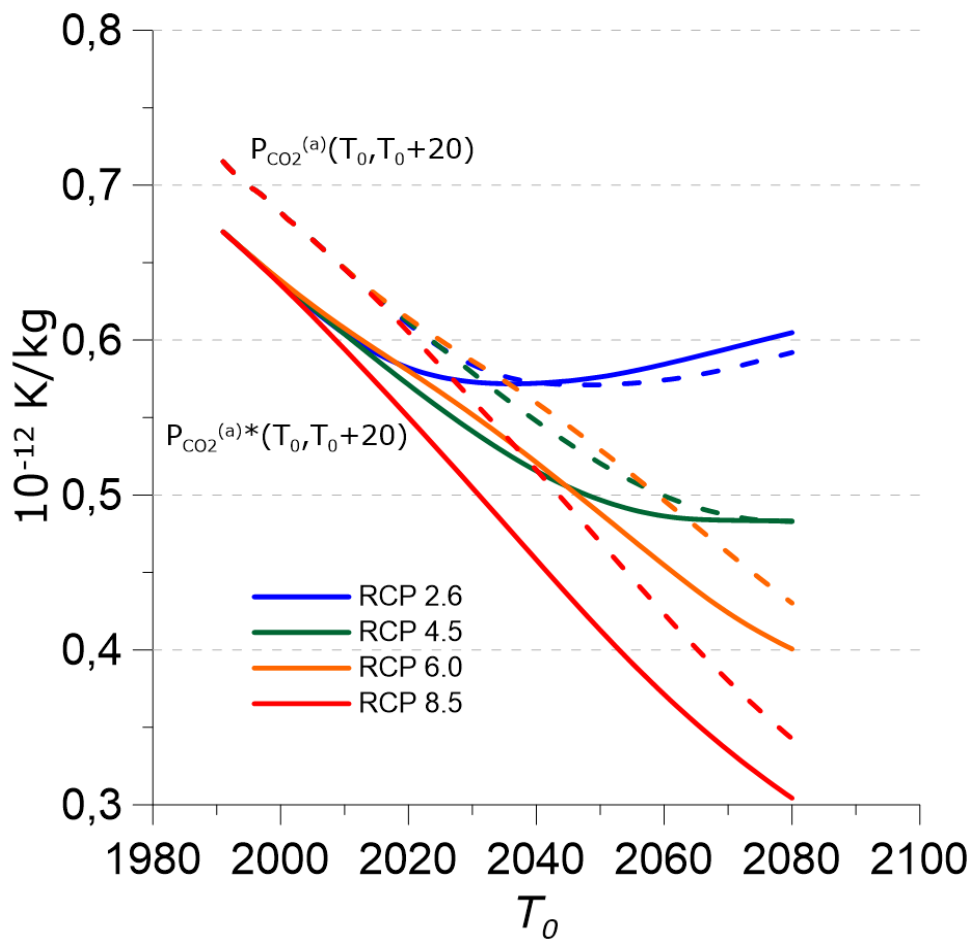

Fig. 3 20-year absolute global temperature change potential for pulse $CO_2$ emissions at time $T_0$ with varying background conditions (solid lines) and with those conditions prescribed according to their values at the starting year (dotted line) varying background conditions.

Fig. 3 shows the 20-year values of the absolute potentials $P^{(a)}$ and $P^{(a)}*$ for $CO_2$. The absolute potential of CO2 under all scenarios except RCP 2.6 decreases throughout the 21st century, and under the most aggressive scenario RCP 8.5 its value falls by more than half. This is due to an increase in the concentration of carbon dioxide in the atmosphere. In the RCP 2.6 scenario, in the second half of the 21st century the atmospheric $CO_2$ concentration starts to decrease, hence the decrease in the potential turns into an increase. Although a relatively short 20-year period after the emission is considered, accounting for changes in the background conditions affects the change in the potential by up to 15%.

The 20-year $P^{(a)}$ and $P^{(a)}*$ for methane behave similarly to those for $CO_2$ (Fig. 4a). A slow decrease in the potentials is replaced by an increase already in the first half of the 21st century for scenario RCP 2.6, and in the second half of the century for RCP 4.5 and 6.0. Accounting for changing background conditions over a 20-year period gives a smaller contribution for the methane potential than for $CO_2$, which does not exceed 3%.

A more rapid decrease in the absolute potentials $P^{(a)}$ and $P^{(a)}*$ for $CO_2$ causes the relative 20-year GTP potential of methane to increase in the 21st century. Excluding changing background conditions, it rises from 70 to 92-108, and from 73 to 91-119 when it is included. In general, we can say that changes in 20-year potentials are delayed by about half the period (7-13 years depending on the potential and scenario of anthropogenic influences) if changing background conditions neglected.

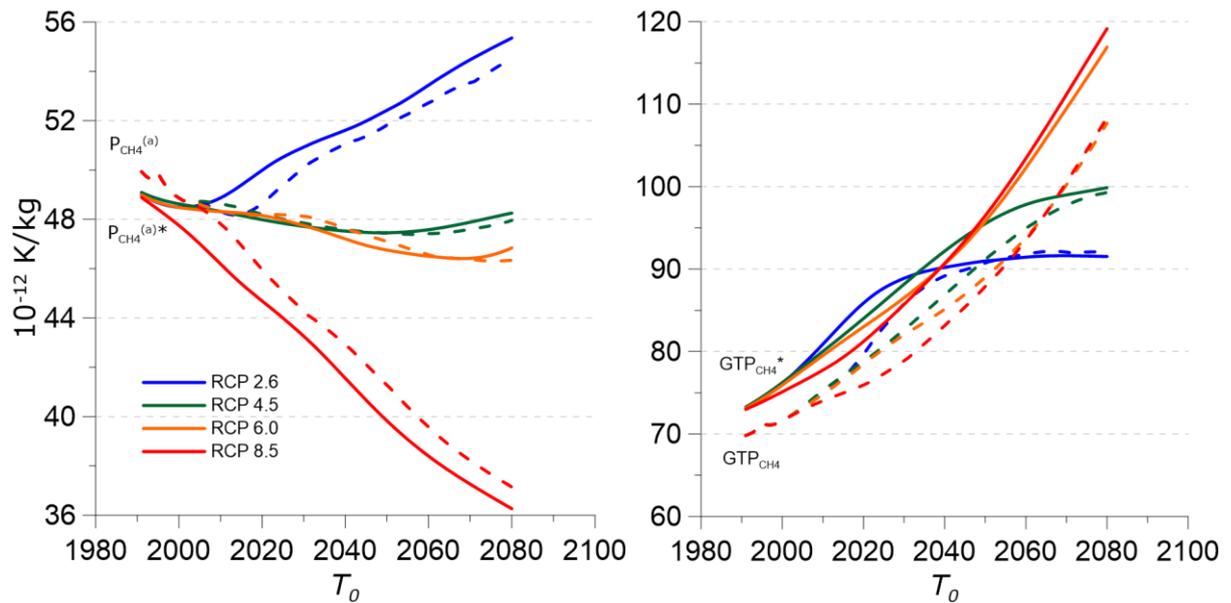

Fig. 4 Similar to Fig. 3, but for pulse CH$_4$ emission (a), relative 20-year CH$_4$ potential (b)

**Results**

**Emissions**

Anthropogenic emissions for Russia were calculated according to the RCP scenarios for the REF region (countries of Eastern Europe and the former USSR, http://www.iiasa.ac.at/web-apps/tnt/RcpDb) by rescaling them to the values of the corresponding emissions from the territory of Russia at the time of the transition from the "Historical simulations" scenario to the RCP scenarios (2000 for methane and 2005 for carbon dioxide). Anthropogenic emissions from the territory of China and North America were calculated similarly according to the RCP scenarios for the ASIA and OECD regions, respectively. Natural emissions were interactively calculated by the IAP RAS CM. It should be noted that according to the data (https://databank.worldbank.org/data/source/world-development-indicators) the reported anthropogenic CH4 emissions from the territory of Russia at the beginning of the 21st century are close to the most aggressive anthropogenic scenario RCP 8.5.

Natural methane emissions from the territory of Russia (Fig. 5a) are estimated to increase by the end of the 21st century by 10-200% depending on the scenario of anthropogenic forcing. Under all scenarios except RCP 8.5, in the second half of the 21st century they reach values of the corresponding anthropogenic emissions at the same year. CO$_2$ uptake by terrestrial ecosystems under all scenarios increases at the beginning of the 21st century. Further, under all scenarios considered, the maximum absorption is reached, which is about 0.4-0.6 PgC/year, after which it begins to decrease, and the more intense the scenario of anthropogenic impacts, the later it happens. The obtained estimates are in good agreement with the estimates (Dolman et al. 2012) for the process models (Table 1) and the estimates (Zhang et al. 2014).

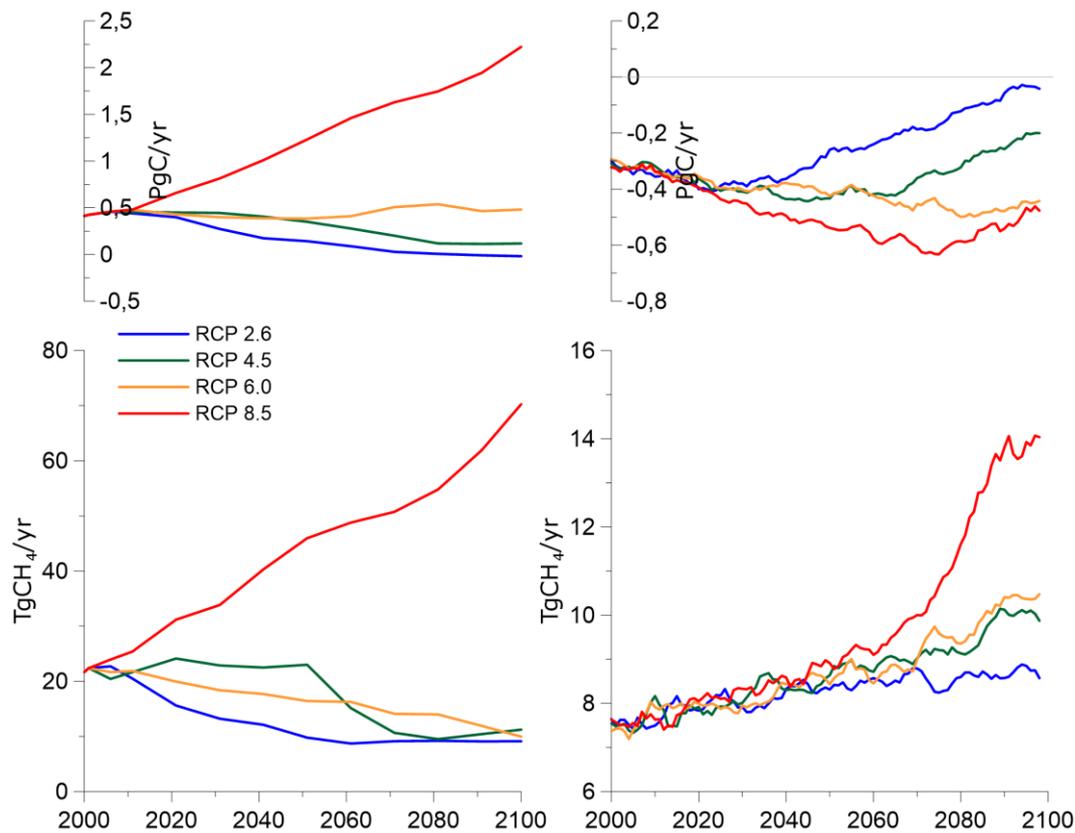

Fig. 5a Anthropogenic (left) and natural (right) emissions of $CO_2$ (top) and $CH_4$ (bottom) from the territory of Russia

| Data | CO2 flux [PgC/yr] | | | |
|---|---|---|---|---|
| | Russia | Canada | USA | China |
| IAP RAS CM (this study, 1995-2005 average) | -0.31 | -0.17 | -0.14 | -0.18 |
| DLEM (Tian et al. 2015) | | -0.15 | -0.34 | |
| (Hayes et al. 2012)* | | -0.24/-0.12/-0.04 | -0.69/-0.36/-0.3 | |
| RCA-GUESS (Zhang et al. 2014)** | -0.32 | | | |
| (Piao et al. 2009)* | | | | -0.35/-0.17/-0.18 |
| (Pacala et al. 2001) | | | -0.3..-0.58 | |
| (Dolman et al. 2012)*** | -0.65/-0.56/-0.76/-0.2 | | | |
| (Tian et al. 2011) | | | | -0.26 |
| (Xiao et al. 2011) | | | -0.64 | |
| | | | | |

* Inverse models/Forward models/Inventory-based estimate
** Arctic tundra only 1990-2100
***Inverse models/Eddy covariance/Land ecosystem accessment/DVGMs

Table 1. Natural $CO_2$ fluxes at the beginning of the 21$^{st}$ century

Natural methane emissions from the territory of China (Fig. 5b) are estimated to increase significantly slower in the 21st century than those from Russia. Only under the most aggressive anthropogenic scenario the increase exceeds 50% by the end of the century. The $CO_2$ flux from terrestrial ecosystems in China into the atmosphere changes in the 21st century in a similar way to the fluxes from the territory of Russia (and all the other regions considered). The uptake maxima in the 21st century amount to 0.2-0.35 PgC/year. The obtained estimates are in general agreement with the estimates (Piao et al. 2009; Tian et al. 2011). It should be noted that for China the values of natural greenhouse gas fluxes in general according to calculations are 5-20% of the corresponding anthropogenic emissions, and China's contribution to the global temperature change is determined by human impact.

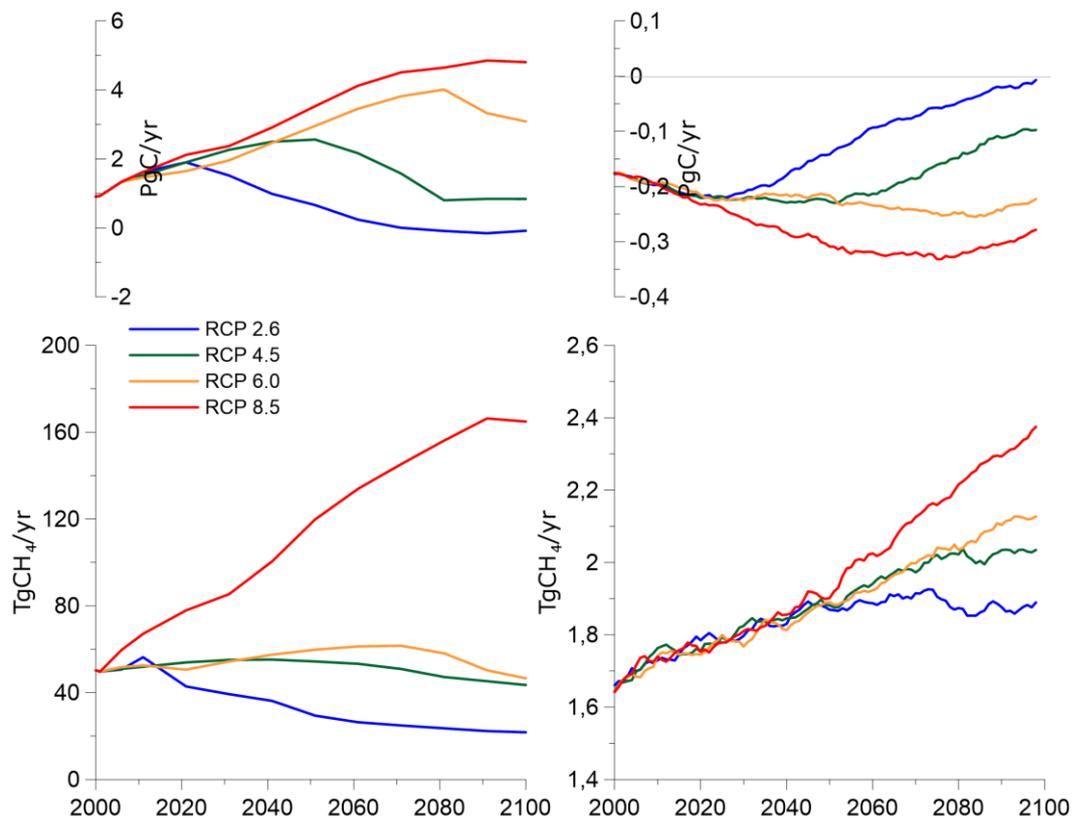

Fig. 5b Anthropogenic (left) and natural (right) emissions of $CO_2$ (top) and $CH_4$ (bottom) from the territory of China

Natural methane emissions from the territory of North America (Fig. 5c) according to calculations in the 21st century are approximately twice as high as emissions from the territory of Russia, but they grow somewhat slower (by 20-100% depending on the scenario). Nevertheless, they begin to exceed the corresponding anthropogenic emissions under all scenarios except RCP 8.5 in the second half of the 21st century. $CO_2$ uptake by terrestrial ecosystems in North America is about 0.3 PgC/yr at the beginning of the 21st century, increases during the century to values of 0.4-0.6 PgC/yr depending on the anthropogenic scenario and then begins to decline. The obtained estimates of natural $CO_2$ fluxes agree well with the data (Tian et al. 2015; Hayes et al. 2012) for the territory of Canada. At the same time, $CO_2$ uptake by terrestrial ecosystems in the United States is somewhat underestimated, both compared to the same data and to the works (Pacala et al. 2001; Xiao et al. 2011). It is worth noting that the major source of the anthropogenic greenhouse gas emissions of North America is the territory of the United States, while the contribution of terrestrial ecosystems of the United States and Canada to $CO_2$ uptake is close to each other, and the main source of natural methane emissions is the territory of Canada (Fig 5d,e).

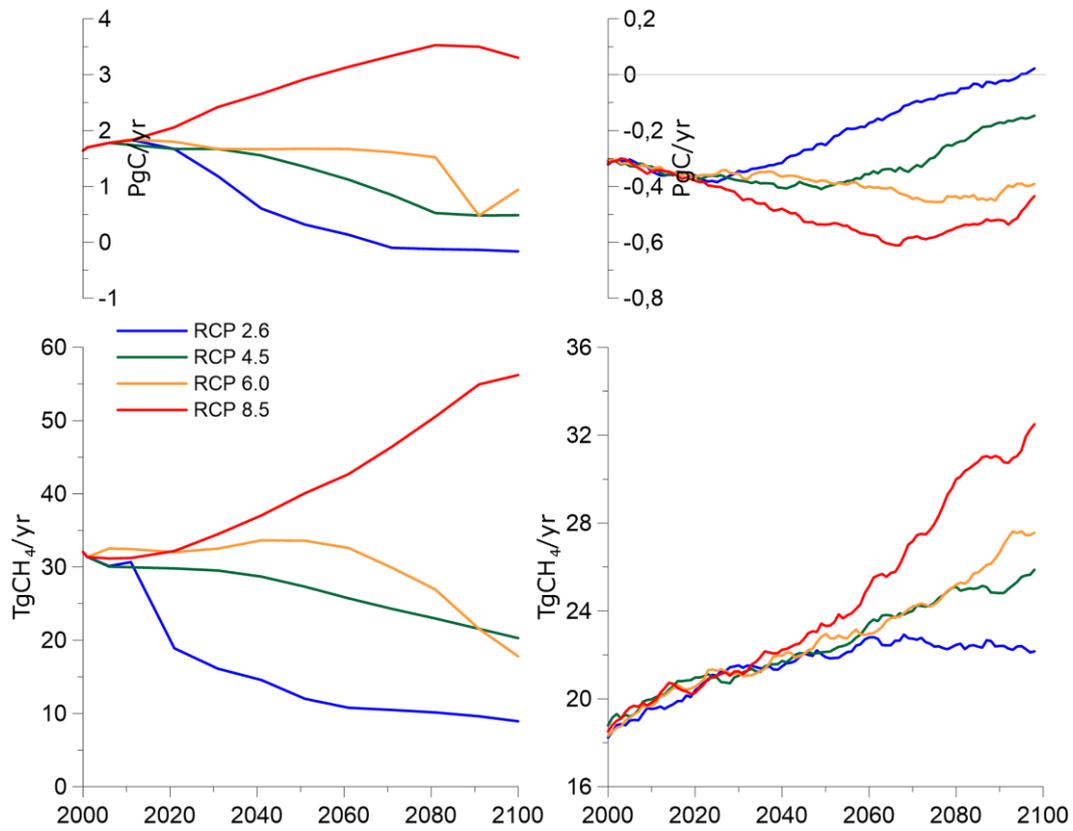

Fig. 5c Anthropogenic (left) and natural (right) emissions of $CO_2$ (top) and $CH_4$ (bottom) from the territory of North America

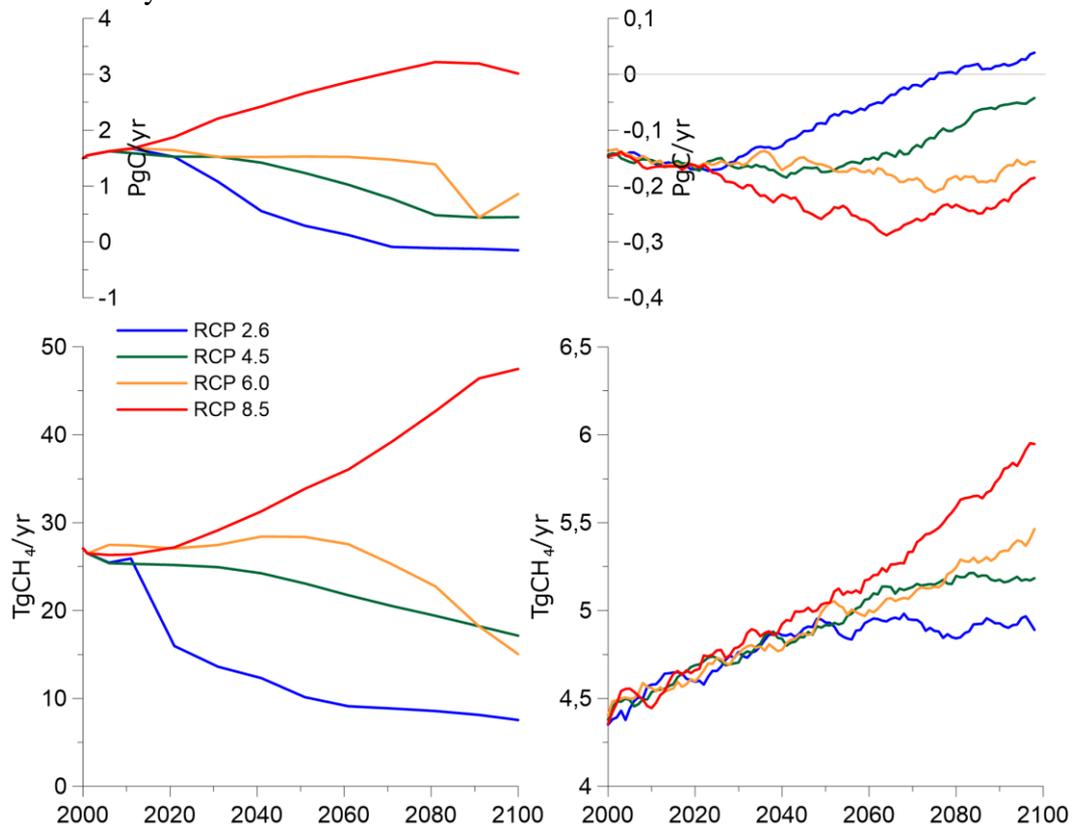

Fig. 5d Anthropogenic (left) and natural (right) emissions of $CO_2$ (top) and $CH_4$ (bottom) from the territory of United States

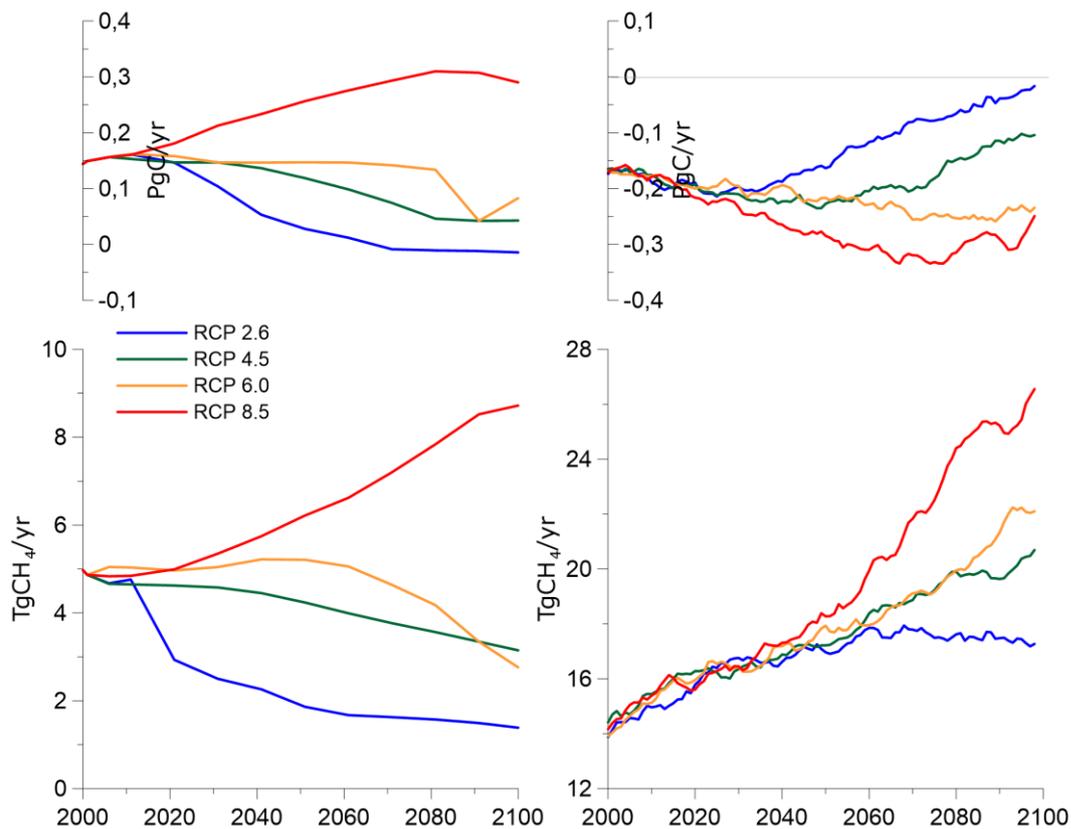

Fig. 5e Anthropogenic (left) and natural (right) emissions of $CO_2$ (top) and $CH_4$ (bottom) from the territory of Canada

**Cumulative temperature potential**

Fig. 6 shows the values of the cumulative CT potential over the interval $[1990;T_H]$ separately for carbon dioxide and methane emissions from Russia, China, and North America (including the United States and Canada separately) with $T_H$ corresponding to the years 2030, 2060, and 2090. It is noted that the anthropogenic CT (Table 2) decreases in the second half of the 21st century under RCP 2.6 (the anthropogenic potential of Russia also decreases under RCP 4.5 and that of North America under RCP 6.0) and increases under scenarios of more intense anthropogenic impacts. The decrease in the anthropogenic potential of Russia occurs mainly as a result of a decrease in anthropogenic methane emissions, while for other regions the contribution of methane and $CO_2$ emission reductions is comparable. For anthropogenic $CO_2$ emissions from the territory of China, it is possible to clearly see the effect of accounting for changes in the background conditions ($CO_2$ radiative forcing according to Formula 12): although throughout the entire 21st century under the RCP 8.5 scenario $CO_2$ emissions are higher than under RCP 6.0, due to the higher $CO_2$ concentration in the atmosphere (and thus a lower forcing) their total impact on the climate by the end of the century is lower. Thus, according to the assessments obtained from the scenarios considered, only RCP 2.6 will lead to the stabilization of the anthropogenic impact on the global atmospheric temperature in the 21st century. At the same time, the stabilizing contribution of natural greenhouse gas fluxes from terrestrial ecosystems in the 21st century increases for Russia, China and the United States under all scenarios considered, and changes little for Canada.

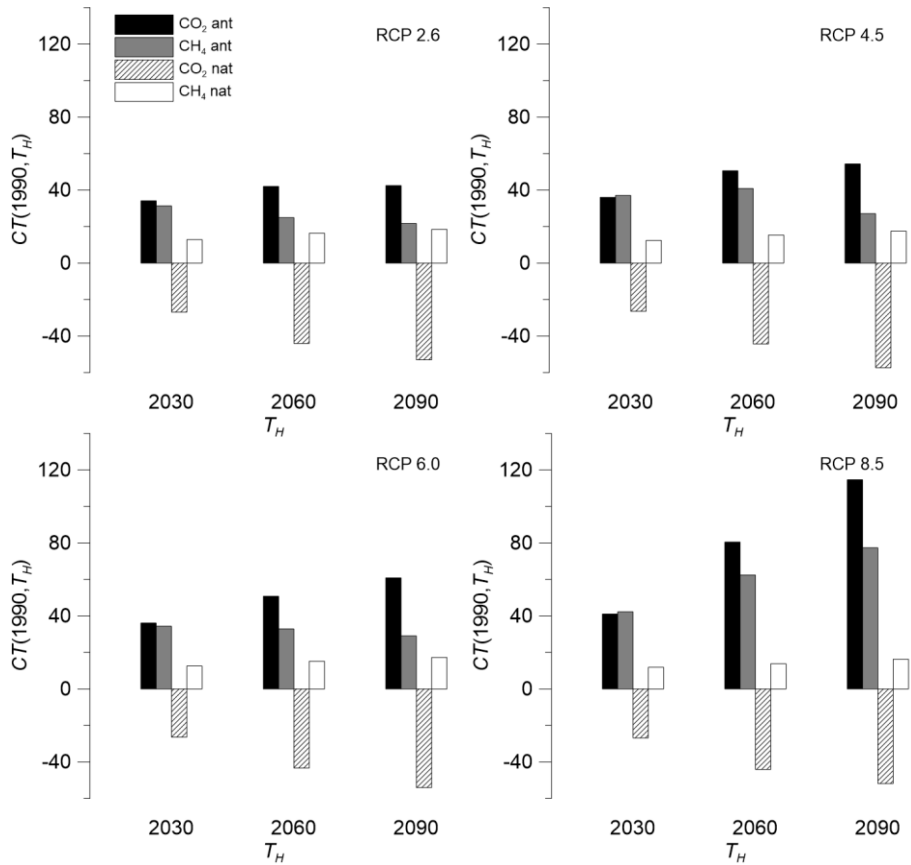

Fig. 6a Absolute potential [mK] of anthropogenic and natural $CO_2$ and $CH_4$ emissions from the territory of Russia for the time intervals [1990;2030], [1990,2060], and [1990;2090].

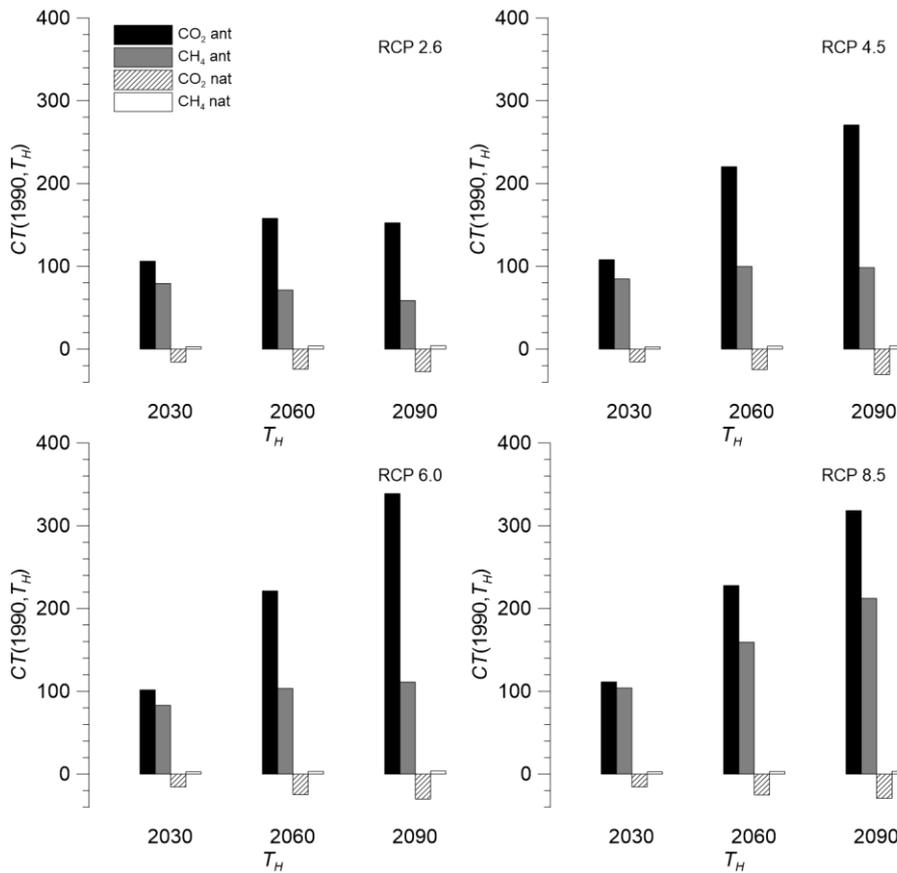

Fig. 6b Absolute potential [mK] of anthropogenic and natural $CO_2$ and $CH_4$ emissions from the territory of China for the time intervals [1990;2030], [1990,2060], and [1990;2090].

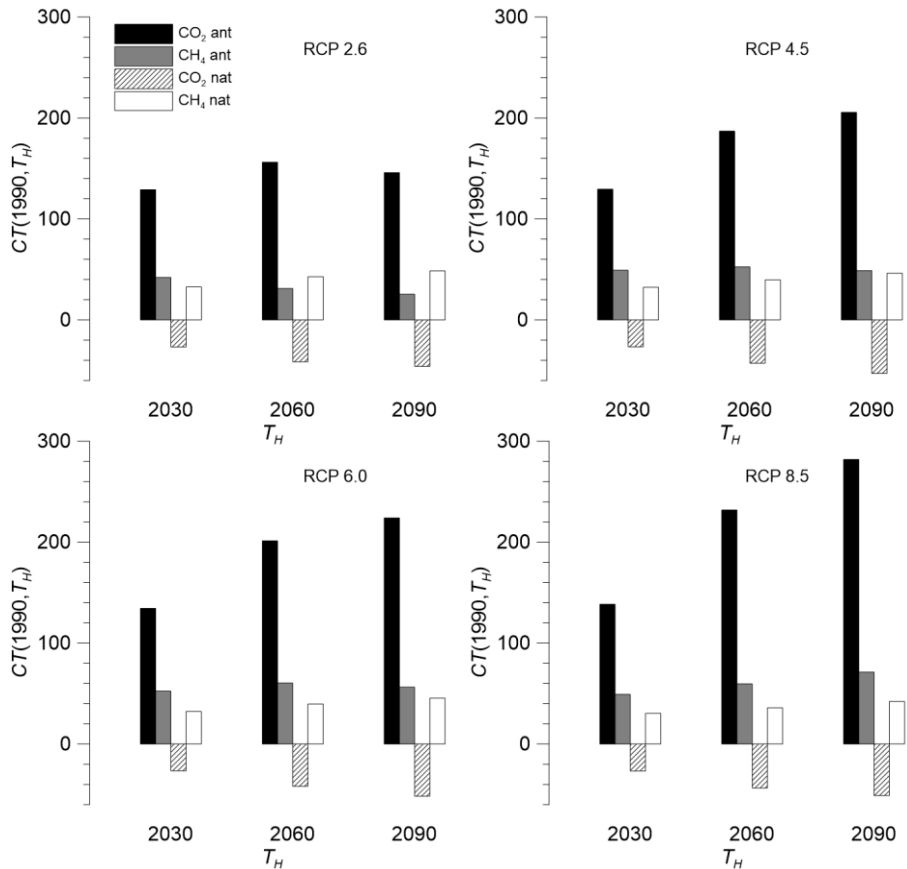

Fig. 6c Absolute potential [mK] of anthropogenic and natural $CO_2$ and $CH_4$ emissions from the territory of North America for the time intervals [1990;2030], [1990,2060], and [1990;2090].

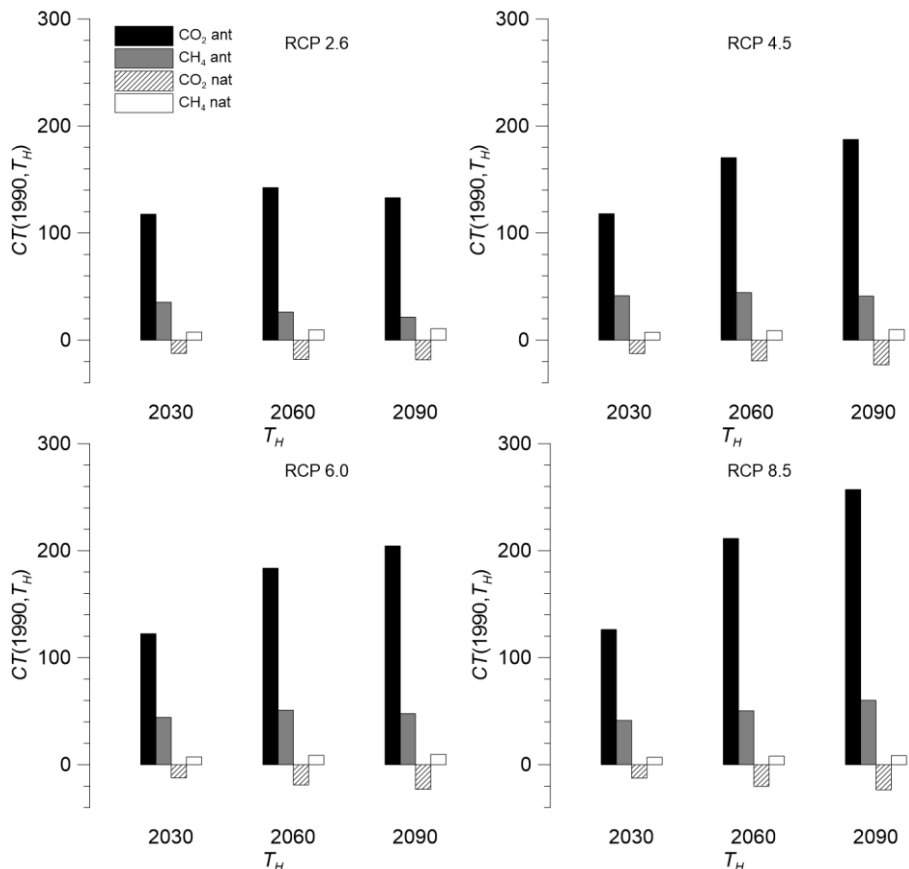

Fig. 6d Absolute potential [mK] of anthropogenic and natural $CO_2$ and $CH_4$ emissions from the territory of United States for the time intervals [1990;2030], [1990,2060], and [1990;2090].

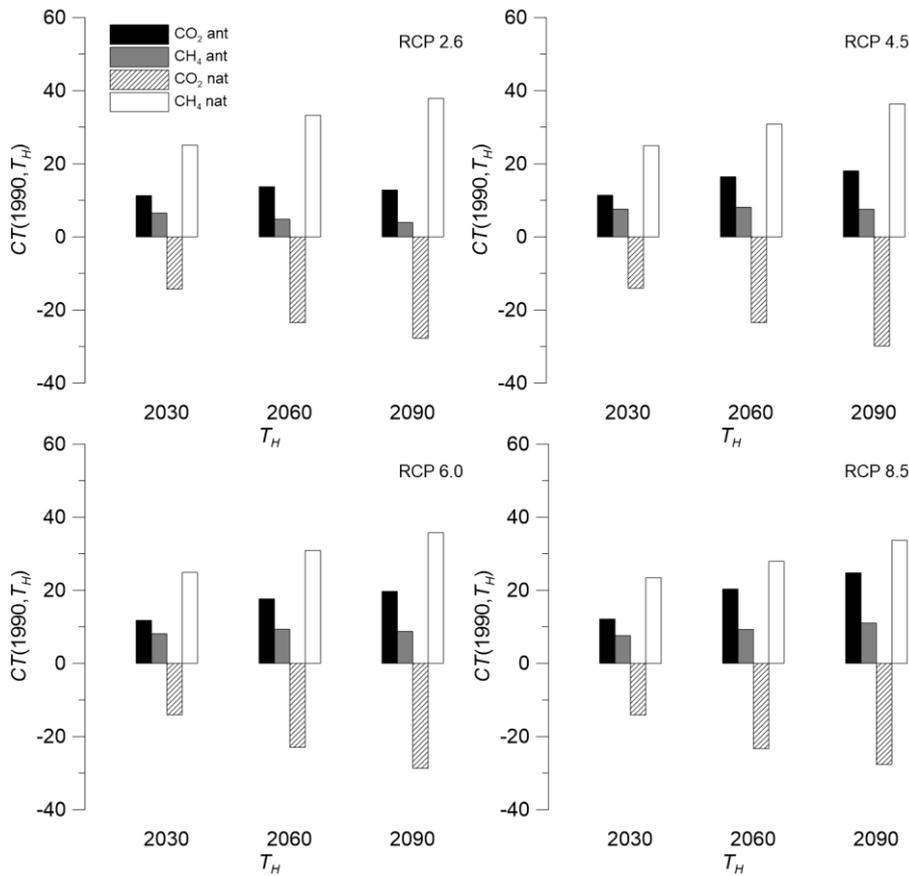

Fig. 6e Absolute potential [mK] of anthropogenic and natural $CO_2$ and $CH_4$ emissions from the territory of Canada for the time intervals [1990;2030], [1990,2060], and [1990;2090].

For Russia, natural greenhouse gas fluxes make a significant contribution to the net impact on global temperature changes, and their stabilizing effect on the climate exceeds the total stabilizing natural impact of other regions. At the same time, for the United States and China, natural emissions are not significant compared to anthropogenic emissions. For North America, it is possible to identify the leading factor - anthropogenic $CO_2$ emissions, which make a greater contribution to the global temperature change than the sum of the moduli of CT of the other fluxes of the greenhouse gases under consideration. In addition, the effect of $CO_2$ uptake by the end of the 21st century begins to exceed the effect of methane emissions under all scenarios except RCP 2.6, and the natural fluxes of greenhouse gases from the territory of North America are calculated to slow warming overall. These effects for North America become even more pronounced when considering the United States and Canada separately. For the United States, anthropogenic $CO_2$ emissions are estimated to be the main factor affecting the climate, compared to which the other greenhouse gas fluxes are insignificant. For Canada, in turn, anthropogenic fluxes are insignificant. According to estimates, the main factors affecting the climate are the $CH_4$ and $CO_2$ fluxes, which are comparable in magnitude, but opposite in direction. Overall, the total contribution to global temperature change from 1990 to the end of the 21st century of anthropogenic and natural fluxes of $CO_2$ and $CH_4$, depending on the scenario of anthropogenic impacts, is 0.03-0.17 K for Russia, 0.18-0.54 K for China and 0.17-0.36 K for North America (including 0.14-0.32 K for the United States and 0.03-0.04 K for Canada), and for Russia it stops amplify warming under all scenarios except RCP 8.5, for China only under RCP 2.6, and for North America under RCP 2.6 and 6.0.

|   |   | RCP 2.6 | | | RCP 4.5 | | | RCP 6.0 | | | RCP 8.5 | | |
|---|---|---|---|---|---|---|---|---|---|---|---|---|---|
|   |   | Anthropogenic emissions | Natural emissions | Net CT | Anthropogenic emissions | Natural emissions | Net CT | Anthropogenic emissions | Natural emissions | Net CT | Anthropogenic emissions | Natural emissions | Net CT |
| Russia | $T_H$=2030 | 65 | -14 | 51 | 73 | -14 | 59 | 70 | -14 | 56 | 83 | -15 | 68 |
|   | $T_H$=2050 | 68 | -24 | 44 | 88 | -25 | 63 | 81 | -24 | 57 | 123 | -26 | 97 |
|   | $T_H$=2075 | 65 | -32 | 33 | 87 | -35 | 52 | 87 | -33 | 54 | 169 | -35 | 134 |
|   | $T_H$=2100 | 64 | -35 | 29 | 80 | -41 | 39 | 91 | -38 | 53 | 208 | -34 | 174 |
| China | $T_H$=2030 | 185 | -13 | 172 | 192 | -13 | 179 | 185 | -13 | 172 | 215 | -13 | 202 |
|   | $T_H$=2050 | 226 | -19 | 207 | 283 | -19 | 264 | 276 | -19 | 257 | 329 | -20 | 309 |
|   | $T_H$=2075 | 222 | -22 | 200 | 356 | -25 | 331 | 395 | -24 | 371 | 466 | -25 | 441 |
|   | $T_H$=2100 | 205 | -23 | 182 | 373 | -28 | 345 | 472 | -27 | 445 | 565 | -26 | 539 |
| North America | $T_H$=2030 | 171 | 6 | 177 | 179 | 6 | 185 | 187 | 6 | 193 | 188 | 3 | 191 |
|   | $T_H$=2050 | 189 | 2 | 191 | 226 | 0 | 226 | 243 | 0 | 243 | 262 | -4 | 258 |
|   | $T_H$=2075 | 180 | 1 | 181 | 250 | -6 | 244 | 279 | -5 | 274 | 326 | -10 | 316 |
|   | $T_H$=2100 | 166 | 4 | 170 | 256 | -6 | 250 | 271 | -6 | 265 | 367 | -7 | 360 |
| United States | $T_H$=2030 | 153 | -5 | 148 | 160 | -5 | 155 | 167 | -5 | 162 | 168 | -6 | 162 |
|   | $T_H$=2050 | 170 | -8 | 162 | 203 | -9 | 192 | 218 | -9 | 209 | 235 | -10 | 225 |
|   | $T_H$=2075 | 162 | -9 | 153 | 225 | -12 | 213 | 250 | -12 | 238 | 293 | -14 | 279 |
|   | $T_H$=2100 | 150 | -7 | 143 | 230 | -13 | 217 | 244 | -14 | 230 | 330 | -15 | 315 |
| Canada | $T_H$=2030 | 18 | 11 | 29 | 19 | 11 | 30 | 20 | 11 | 31 | 20 | 9 | 29 |
|   | $T_H$=2050 | 19 | 10 | 29 | 23 | 9 | 32 | 25 | 9 | 34 | 27 | 6 | 33 |
|   | $T_H$=2075 | 18 | 10 | 28 | 25 | 6 | 31 | 29 | 7 | 36 | 33 | 4 | 37 |
|   | $T_H$=2100 | 16 | 11 | 27 | 26 | 7 | 33 | 27 | 8 | 35 | 37 | 8 | 45 |

Table 2. Anthropogenic and natural CT(1990,$T_H$) [mK]

**Conclusions**

According to the results obtained, accounting for changes in climatic conditions can strongly influence the impact indicators of various greenhouse gas emissions on the climate system, especially over large time horizons under the most aggressive scenarios of anthropogenic impact. The more intensive the anthropogenic impact on the climate, the smaller the 100-year potential of contemporary $CO_2$ emissions is. At the same time, the relative role of contemporary methane emissions increases.

We acknowledge that some processes are missed in our assessment:
- Release of methane and nitrogen dioxide from natural fires (Saunois et al. 2020);
- $CO_2$ and $CH_4$ release from permafrost thaw which accompanies climate warming (Schuur et al. 2015; SROCL);
- $CH_4$ and $N_2O$ release from inland water bodies (Deemer et al. 2016; Stepanenko et al. 2011);
- Complete neglect of $N_2O$ release from the terrestrial ecosystems (Tian et al. 2015; SROCL).

Natural fires release 15±3 Tg of methane and 0.9±3 Tg of nitrogen dioxide per year with ~10% contribution from the Russian territory (Eliseev and Vasileva 2020). These values are sufficiently small in order not to change the bottom right panels in Figs.5a-d. However, they add to turning the specific area to the source of the greenhouse gases (say, in CO2-equivalents).

Release of greenhouse gases from inland water bodies is very uncertain (Deemer et al., 2016) but again may add to turning the specific area to the greenhouse-gases source.

In addition, the quantitative assessments of this paper may be refined by taking into account the effect of $CO_2$ and $CH_4$ release from soil into the atmosphere during thawing of permafrost soils of subpolar latitudes, of which "old" (formed in the last interglacials, not decomposed due to cold conditions and not considered by the contemporary IAP RAS CM carbon and methane cycle blocks) carbon substrate under scenario RCP 8.5 in the 21st century could release up to 174 Pg of carbon in the form of carbon dioxide and methane (Schuur et al. 2015). In addition, the development of thermokarst lakes (also not considered in this paper) during the thawing of permafrost terrestrial regions contributes to the development of lake taliks, which also contributes to the release of these greenhouse gases into the atmosphere - according to available estimates in the 21st century under RCP scenarios this could lead to the release of up to 50 Tg $CH_4$ into the atmosphere with the largest contribution of the first half of the century (Schneider von Deimling et al. 2015). It should also be kept in mind that the trajectory of the response of the characteristics of the terrestrial carbon and methane cycles to climatic changes essentially depends on the interaction with the nitrogen cycle (Myhre et al. 2013; SROCL; Schuur et al. 2015).

When making decisions, it should be taken into account that the role of natural fluxes of greenhouse gases into the atmosphere from terrestrial ecosystems is changing depending on the planning horizon. For all the regions considered, the uptake of $CO_2$ by terrestrial ecosystems in all scenarios begins to decrease in the second half of the 21st century. Accordingly, the stabilizing effect of $CO_2$ fluxes may gradually lose importance. At the same time, methane emissions in all of the considered regions are increasing significantly. According to the results obtained under the RCP 2.6 scenario of anthropogenic emissions in climatic conditions of the second half of the 21st century ($T_0 > 2050$), natural greenhouse gas fluxes in considered regions will accelerate climate warming.

**Acknoledgements:**

This study was supported by the Russian Science Foundation (project 19-17-00240). The analysis of the regional features in Eurasia was carried out within the framework of grant agreement (no. 075-15-2020-776) with the Ministry of Science and Higher Education of the Russian Federation. Estimates of the natural carbon and methane fluxes for different time horizons were obtained in a framework of the Russian Science Foundation project no. 21-17-00012.

**Supplementary material**

Absolute Global Temperature change Potential (P[(a)]) is the change in global mean surface temperature at time H in response to a 1 kg pulse emission of gas x at time t = 0. It is often written as a convolution of the radiative forcing with the climatic response kernel $R_T$:

$$P_x^{(a)}(H) = \int_0^H RF_x(t) R_T(H-t) dt, \qquad (1)$$

where $RF_x$ is the radiative forcing due to a pulse emission of a gas x, and $R_T$ is the temporally displaced climate response to a unit forcing. $RF_x$, can be written as the product of gas x radiative efficiency, $A_x$, and the perturbation in its burden, $IRF_x$.

**Radiative forcing of $CO_2$**

For carbon dioxide radiative forcing can be written as:

$$RF_{CO_2}(t) = A_{CO_2} IRF_{CO_2}(t), \qquad (2)$$

where $A_{CO2}$ is the radiative forcing per unit mass increase in atmospheric abundance of $CO_2$, and $IRF_{CO2}(t)$ is the impulse response function (or Green's function).
For sufficiently small emissions and approximately constant background conditions the radiative forcing for $CO_2$ can be approximated as time-invariant using the expression based on radiative transfer models (Myhre et al. 1998):

$$A_{CO_2} = \frac{\alpha \ln\left(\frac{C_0 + \Delta C}{C_0}\right)}{\Delta C} \left(\frac{M_a}{M_{CO_2}}\right) \left(\frac{10^9}{T_m}\right), \qquad (3)$$

where α = 5.35 W m$^{-2}$, $C_0$ is the reference concentration and ΔC is the change from this reference. The mean molecular weight of air $M_a$ (28.97 kg kmol$^{-1}$), the molecular weight of $CO_2$ $M_{CO2}$ and the total mass of the atmosphere $T_m$ (5.1352 ×10$^{18}$ kg (Trenberth and Smith 2005)) used to convert it to per kg value (Shine et al. 2005). $A_{CO2}$ equals 1.7517*10$^{-15}$ W m$^{-2}$ kg$^{-1}$ at 391 ppm $CO_2$ levels.

| $i$ | 0 | 1 | 2 | 3 |
|---|---|---|---|---|
| $a_i$ [unitless] | 0.2173 | 0.2240 | 0.2842 | 0.2763 |
| $\tau_i$ [years] | - | 394.4 | 36.54 | 4.304 |

Table 1. Equation 4 parameters.

$IRF_{CO2}$ cannot be represented by a simple exponential decay. The decay of a perturbation of atmospheric $CO_2$ following a pulse emission at time t is usually approximated by a sum of exponentials (Joos et al. 2013) with parameters listed in Table 1:

$$IRF_{CO_2}(t) = a_0 + \sum_{i=1}^{N} a_i \exp\left(-\frac{t}{\tau_i}\right), \qquad (4)$$

**Radiative forcing of $CH_4$**

For methane radiative forcing can be written (similarly to (2)) as:

$$RF_{CH_4}(t) = (1 + f_1 + f_2) A_{CH_4} IRF_{CH_4}(t), \qquad (5)$$

where $f_1 = 0.5$ and $f_2 = 0.15$ are used to include effects of methane on ozone and stratospheric H$_2$O correspondingly (Myhre et al. 2013). The radiative forcing for CH$_4$ can be approximated following (Myhre et al. 1998):

$$A_{CH_4} = \frac{\left(\beta(\sqrt{Mt_0+\Delta Mt}-\sqrt{Mt_0})-(f(Mt_0+\Delta Mt,N)-f(Mt_0,N))\right)}{\Delta Mt}\left(\frac{M_a}{M_{CH_4}}\right)\left(\frac{10^9}{T_m}\right), \quad (6)$$

$$f(Mt,N) = 0.47\ ln(1 + 2.01 \times 10^{-5}(MtN)^{0.75} + 5.31 \times 10^{-15}Mt(MtN)^{1.52}), \quad (7)$$

where β = 0.036 W m$^{-2}$, Mt$_0$ [ppb] is the reference concentration of CH$_4$ and N [ppb] is the concentration of N$_2$O. At current levels (1803 ppb CH$_4$ and 324 ppb N$_2$O (Myhre et al. 2013)) A$_{CH4}$ equals 1.28*10$^{-13}$ W m$^{-2}$ kg$^{-1}$. IRF$_{CH4}$ can be approximated by a simple exponential decay:

$$IRF_{CH_4}(t) = exp\left(-\frac{t}{\tau_{CH_4}}\right), \quad (8)$$

where $\tau_{CH4}$ is the methane perturbation lifetime.

**Methane perturbation lifetime**

Methane perturbation lifetime in atmosphere, taking into account main sinks (listed in Table 2), can be written as:

$$\tau_{CH_4} = 1.34 \frac{1}{\sum_{l=1}^{4}\frac{1}{mlt_l}} \quad (9)$$

| $l$ | Methane sinks | $mlt_l$ [years] |
|---|---|---|
| 1 | Tropospheric OH | $\tau_{OH}(t)$ |
| 2 | Soil | 150 |
| 3 | Stratosphere | 120 |
| 4 | Tropospheric chlorine | 200 |

Table 2. Equation 9 parameters.

Methane lifetime in reaction with tropospheric OH was calculated based on (Holmes et al., 2013) parametrization using individual sensitivity coefficients s$_i$ to main factors F$_i$ (Table 3):

$$s_i = \frac{d\ ln(\tau_{OH})}{d\ ln(F_i)} \quad (10)$$

Sensitivity parameters and the time series of corresponding factors let us to build a parametric model for methane lifetime approximation:

$$ln(\tau_{OH}(t)) = ln(\langle\tau_{OH}\rangle) + \sum_i s_i \Delta\ ln(F_i(t)) \quad (11)$$
$$ln\left(\frac{\tau_{OH}(t)}{\langle\tau_{OH}\rangle}\right) = \sum_i s_i\ ln\left(\frac{F_i(t)}{F_{0,i}}\right) \quad (12)$$
$$\tau_{OH}(t) = \tau_0\left(1 + \sum_i s_i\left(\frac{F_i(t)-F_{0,i}}{F_{0,i}}\right)\right) \quad (13)$$

| $i$ | Variable | Sensitivity ($s_i$) |
|---|---|---|
| Chemistry-climate interactions | | |
| 1 | Air Temperature | -3.0 |
| 2 | Biomass Burning emissions | +0.02 |
| 3 | CH4 abundance | +0.31 |
| Anthropogenic emissions | | |
| 4 | Land NOx | -0.14 |
| 5 | CO | +0.06 |
| 6 | VOC | +0.04 |

Table 3. Sensitivity coefficients for equations 10-13 (Holmes et al., 2013).

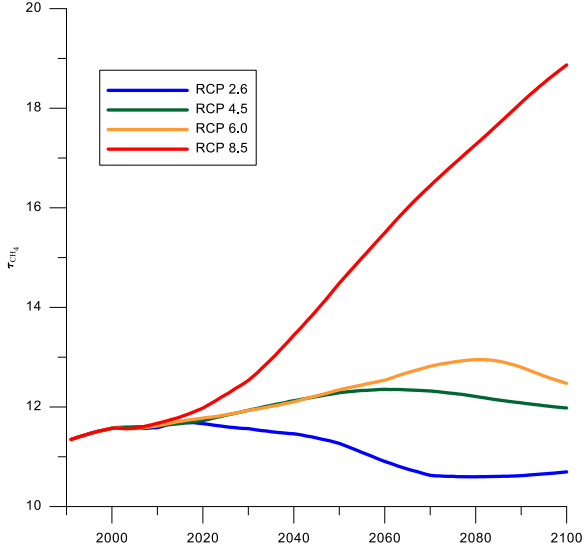

Fig. 1 Calculated $\tau_{OH}(t)$ [yr] under different scenarios of anthropogenic impact.

**Climate response function**

Climate response function $R_T$ can be represented as a sum of exponentials:

$$R_T(t) = \sum_{j=1}^{M} \frac{c_j}{d_j} exp\left(-\frac{t}{d_j}\right), \quad (14)$$

| $j$ | 1 | 2 |
|---|---|---|
| $c_j$ [K(W m$^{-2}$)$^{-1}$] | 0.631 | 0.429 |
| $d_j$ [years] | 8.4 | 409.5 |

Table 4. Equation 14 parameters.

where the parameters $c_j$ are the components of the climate sensitivity and $d_j$ are corresponding response times (Table 2) (Myhre et al. 2013).

**Modified absolute Global Temperature change Potential**

For changing background conditions P$^{(a)}$ can be rewritten as sum of integrals for each year:

$$P_x^{(a)*}(T_0, T_H) = \sum_{k=T_0+1}^{T_H} \int_{k-1}^{k} RF_{x,k}(t) R_T(T_H - T_0 - t) dt, \quad (15)$$

where $T_0$ is the year of emission and $T_H = T_0 + H$. $RF_{x,k}$ can be achieved in assumption of all required parameters being step functions, constant for each particular year k

Equations 2,3,5,6 and 8 can be rewritten as:

$$RF_{CO_2,k}(t) = A_{CO_2,k} IRF_{CO_2}(t) \tag{16}$$

$$A_{CO_2,k} = \frac{\alpha}{C_k}\left(\frac{M_a}{M_{CO_2}}\right)\left(\frac{10^9}{T_m}\right) \tag{17}$$

$$RF_{CH_4,k}(t) = (1 + f_1 + f_2) A_{CH_4,k} IRF_{CH_4,k}(t) \tag{18}$$

$$A_{CH_4,k} = \left(\frac{\beta}{2\sqrt{Mt_k}} - \frac{0.47\left(1.5075 \times 10^{-5} Mt_k^{-0.25} N_k^{0.75} + 13.3812 \times 10^{-15} (Mt_k N_k)^{1.52}\right)}{1 + 2.01 \times 10^{-5} (Mt_k N_k)^{0.75} + 5.31 \times 10^{-15} Mt_k (Mt_k N_k)^{1.52}}\right)\left(\frac{M_a}{M_{CH_4}}\right)\left(\frac{10^9}{T_m}\right) \tag{19}$$

$$IRF_{CH_4,k}(t) = exp\left(-\frac{t}{\tau_{CH_4,k}}\right) \tag{20}$$